\documentclass[11pt]{article}

\usepackage{graphicx}
\usepackage{amsmath}
\usepackage{dcolumn}
\usepackage{bm}
\usepackage{cite}
\usepackage[utf8]{inputenc}
\usepackage{authblk}

\newcommand{\be}{\begin{equation}}
\newcommand{\ee}{\end{equation}}
\newcommand{\ba}{\begin{eqnarray}}
\newcommand{\ea}{\end{eqnarray}}
\newcommand{\no}{\nonumber \\}
\newcommand{\gsim}{\mathrel{\hbox{\rlap{\lower.55ex \hbox {$\sim$}}
                   \kern-.3em \raise.4ex \hbox{$>$}}}}
\newcommand{\lsim}{\mathrel{\hbox{\rlap{\lower.55ex \hbox {$\sim$}}
                   \kern-.3em \raise.4ex \hbox{$<$}}}}

\def\roughly#1{\mathrel{\raise.3ex\hbox{$#1$\kern-.75em%
\lower1ex\hbox{$\sim$}}}}
\def\lsim{\roughly<}
\def\gsim{\roughly>}

\def\({\left(}
\def\){\right)}
\def\[{\left[}
\def\]{\right]}
\def\<{\langle}
\def\>{\rangle}
\def\pd{\partial}

\def\k{{\kappa}}

\def\d{{\delta}}

\def\o{{\omega}}
\def\O{{\Omega}}
\def\e{{\epsilon}}

\def\a{{\alpha}}

\def\c{{\chi}}

\def\g{{\gamma}}
\def\G{{\Gamma}}

\def\m{{\mu}}
\def\n{{\nu}}
\def\r{{\rho}}
\def\s{{\sigma}}

\def\t{{\tau}}
\def\th{{\theta}}
\def\ph{{\phi}}

\def\x{{\xi}}

\newcommand{\Omg}{\Omega}

\newcommand{\wg}{{\wedge}}

\newcommand{\cN}{{\cal N}}
\newcommand{\cG}{{\cal G}}
\newcommand{\tr}{\text{tr}}

\title{{\bf Mass Effect on Axial Charge Dynamics}}

\author[1,3]{Er-dong Guo\thanks{guoerdong@itp.ac.cn}}
\author[2]{Shu Lin\thanks{linshu8@mail.sysu.edu.cn}}
\affil[1]{State Key Laboratory of Theoretical Physics, Institute of
Theoretical Physics, Chinese Academy of Sciences, Beijing 100190, China}
\affil[2]{Institute of Astronomy and Space Sciences, Sun Yat-Sen University, No 135 Xingang Xi Rd, Guangzhou 510275, China}
\affil[3]{Kavli Institute of Theoretical Physics China, Chinese
  Academy of Sciences, Beijing 100190, China}

\date{\today}

\begin{document}

\maketitle

\begin{abstract}
We studied effect of finite quark mass on the dynamics of axial charge using the D3/D7 model in holography. The mass term in axial anomaly equation affects both the fluctuation (generation) and dissipation of axial charge. We studied the dependence of the effect on quark mass and external magnetic field. For axial charge generation, we calculated the mass diffusion rate, which characterizes the helicity flipping rate. The rate is a non-monotonous function of mass and can be significantly enhanced by the magnetic field. The diffusive behavior is also related to a divergent susceptibility of axial charge. For axial charge dissipation, we found that in the long time limit, the mass term dissipates all the charge effectively generated by parallel electric and magnetic fields. The result is consistent with a relaxation time approximation. The rate of dissipation through mass term is a monotonous increasing function of both quark mass and magnetic field. 
\end{abstract}

\newpage

\section{Introduction}

It is believed that parity odd domains with chiral imbalance are produced in finite
temperature quark-gluon plasma (QGP). Their presence can be
detected via axial anomaly as chiral magnetic effect (CME) \cite{Kharzeev:2007jp,Kharzeev:2007tn,Kharzeev:2004ey,Fukushima:2008xe}
and chiral magnetic wave (CMW) \cite{Kharzeev:2010gd,Burnier:2011bf} in heavy ion collisions, see \cite{Kharzeev:2015znc,Liao:2014ava,Huang:2015oca} for recent reviews. The
former leads to the generation of vector current along the
direction of external magnetic field:
\begin{align}
\vec{j}_V=\frac{N_ce}{2\pi^2}\m_5 \vec{B},
\end{align}
where $\m_5$ is the axial chemical potential characterizing the chiral
imbalance.
The latter leads to the propagation of axial
and vector charges along the direction of external magnetic
field. Analogous effects exist when the magnetic field is replaced by
vorticity of QGP \cite{Son:2009tf,Jiang:2015cva}.
These effects are being intensively searched for in heavy ion collision
experiments in recent years \cite{Adamczyk:2014mzf,Abelev:2009ac,Abelev:2012pa}.

Theoretical descriptions of CME and CMW have been developed in
different frameworks including hydrodynamics \cite{Son:2009tf,Kharzeev:2011ds,Megias:2013joa,Jensen:2013vta,Sadofyev:2010pr} and kinetic theory
\cite{Pu:2010as,Chen:2012ca,Stephanov:2012ki,Stephanov:2014dma,Son:2012wh,Son:2012zy,Wu:2016dam}
etc. Most frameworks assume quarks
being massless, see exception for example in \cite{Chen:2013iga,Kirilin:2013fqa}.
While it is known that finite quark mass does not modify CME, we do
expect quark mass to have imprints on the dynamics of
axial charge. Naively, if
the mass of one quark flavor is much larger than the temperature
of QGP, that quark flavor decouples from axial current. We would like to ask quantitative questions on the mass effect
on dynamics of axial charge. This is relevant in reality because the mass
of strange quark is comparable to the temperature of QGP created at
relativistic heavy ion collider (RHIC) and large hadron collider (LHC).
With the inclusion of mass term, the axial anomaly equation reads
\begin{align}\label{anomaly_m}
\pd_\m j_5^\m=2im\bar{\psi}\g^5\psi-\frac{e^2}{16\pi^2}\e^{\m\n\r\s}F_{\m\n}F_{\r\s}-\frac{g^2}{16\pi^2}\tr \e^{\m\n\r\s}G_{\m\n}G_{\r\s},
\end{align}
where the three terms on the right hand side (RHS) corresponds to mass term, QED anomaly term
and QCD anomaly term respectively. \eqref{anomaly_m} is written for one flavor of quark with mass $m$. All three terms lead to modification
of axial charge dynamics. The effect of QED anomaly term is
extensively studied in the above mentioned references. The effect of
QCD anomaly was studied recently \cite{Fukushima:2010vw,Jimenez-Alba:2014iia,Iatrakis:2014dka,Iatrakis:2015fma}. In this work, we will focus on the effect of the mass term.
On one hand, finite quark mass explicitly breaks axial symmetry, offering a mechanism of axial charge generation. We find that
the mass operator diffuses at low frequency the same way as the Chern-Simon (CS)
number. The diffusion of the CS number is known to generate axial
charge. The same is true for the mass operator. We calculate the
diffusion rate of mass term as a measure of axial charge generation. We also define a dynamical susceptibility by CME, and find it to be divergent in the low frequency limit. We explain the common physical reason for the diffusive mass operator and the divergent susceptibility.
On the other hand, finite quark mass also leads to axial
charge dissipation. The
dissipation effect is studied recently in \cite{Jimenez-Alba:2015awa,Sun:2016gpy} in a relaxation time
approximation. We
will discuss axial charge dissipation in an indirect way: we set
up parallel electric and magnetic field and measure the rate of dissipation through the mass term. 
The situation is further complicated by the existence of a reservoir of adjoint matter, to which axial charge can dissipate. By taking into account the additional loss rate, we find that the axial charge dissipates entirely in the long time limit, which is consistent with the relaxation time approximation.
We will study these effects as a function of both quark mass and external
magnetic field using a holographic model.

The paper is organized as follows: In Sec II we give a self-contained review
of the holographic model. In Sec III we discuss separately mass effect on axial
charge generation and dissipation, which we coined mass diffusion rate and mass dissipation effect respectively. We summarize the results in Sec
IV. We collect technical details in obtaining phase diagram and
hydrodynamic solutions in two appendices.

\section{A quick review of the model}

\subsection{The D3/D7 background}
\begin{figure}[t]
\includegraphics[width=0.5\textwidth]{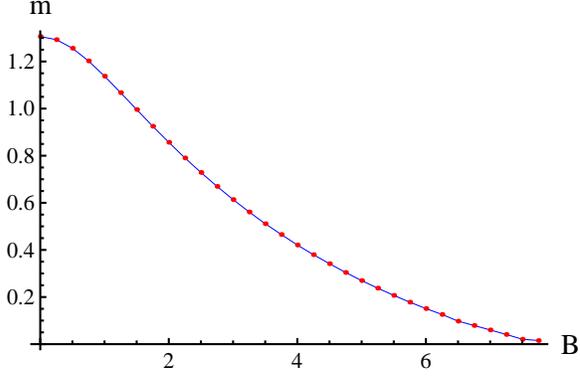}
\caption{\label{fig_mB}$m$-$B$ phase diagram of $D3/D7$ background. The axis labels are dimensionless numbers with units set by $\pi T=1$. The region with small $m$ and $B$ corresponds to the meson melting phase, while the region with large $m$ and $B$ corresponds to the mesonic phase.}
\end{figure}

We use the D3/D7 model to study the effect of finite quark mass. The
background is sourced by $N_c$ D3 branes. The worldvolume fields of D3
branes are ${\cal N}=4$ supersymmetric Yang-Mills (SYM) theory. In addition, there
are $N_f$ D7 branes in the background. The open string
stretching between D3 and D7 branes is dual to ${\cal N}=2$
hypermultiplet.
The ${\cal N}=4$ and ${\cal N}=2$ fields are in the adjoint and fundamental
representations of the $SU(N_c)$ group respectively.
By analogy with
QCD, we will loosely refer to the ${\cal N}=4$ and ${\cal N}=2$ fields as
gluons and quarks respectively. A detailed account of field content can be found
in \cite{Hoyos:2011us}. The ${\cal N}=4$ theory has a $SO(6)_R$ global symmetry,
which is broken by the ${\cal N}=2$ theory to $SO(4)\times U(1)_R$. As
we will see, the $U(1)_R$ symmetry is anomalous. We will identify it
with axial symmetry.
We start with the finite temperature black hole background of D3
branes following the notations of \cite{Mateos:2006nu}:
\begin{align}\label{d3_metric}
ds^2&=g_{tt}dt^2+g_{xx}d\vec{x}^2+g_{\r\r}d\r^2+g_{\th\th}d\th^2+g_{\ph\ph}d\ph^2+g_{SS}d\O_3^2, \no
&=-\frac{r_0^2}{2}\frac{f^2}{H}\r^2dt^2+\frac{r_0^2}{2}H\r^2dx^2+\frac{d\r^2}{\r^2}+d\th^2+\sin^2\th d\ph^2+\cos^2\th d\O_3^2.
\end{align}
where
\begin{align}
f=1-\frac{1}{\r^4},\quad H=1+\frac{1}{\r^4}.
\end{align}
The temperature is fixed by $T=r_0/\pi$.
Note that we have factorized $S_5$ into $S_3$ and two
additional angular coordinates $\th$ and $\ph$, which makes the
breaking of global symmetry $SO(6)_R\to SO(4)\times U(1)_R$ manifest.
There is a nontrivial background Ramond-Ramond form
\begin{align}
C_4=\(\frac{r_0^2}{2}\r^2H\)^2dt\wg dx_1\wg dx_2\wg dx_3-\cos^4\th d\ph\wg d\O_3.
\end{align}
In the probe limit $N_f/N_c\ll 1$, the D7 branes do not backreact on the background of
the D3 branes. This corresponds to the quenched limit of QCD.
The D3 and D7 branes occupy the following dimensions.
\be
\label{table:d3d7}
\begin{array}{c|cccccccccc}
   & x_0 & x_1 & x_2 & x_3 & x_4 & x_5 & x_6 & x_7 & x_8 & x_9\\ \hline
\mbox{D3} & \times & \times & \times & \times & & &  &  & & \\
\mbox{D7} & \times & \times & \times & \times & \times  & \times
& \times & \times &  &   \\
\end{array}
\ee
The D3 and D7 branes are separated in the $x_8$-$x_9$ plane. Using translation symmetry, we put D3 branes at the origin of the plane and parameterize the position of D7 branes by radius $\r\cos\th$ and polar angle $\ph$. The D7 branes have rotational symmetry in the $x_8$-$x_9$ plane, corresponding to $U(1)_R$ symmetry in the dual field theory. We use the symmetry to choose $\ph=0$.
The embedding function $\th(\r)$ of D7 branes in D3 background is
determined by minimizing the action including a DBI term and WZ term
\begin{align}\label{S_bare}
&S_{D7}=S_{DBI}+S_{WZ}, \no
&S_{DBI}=-N_fT_{D7}\int d^8\x\sqrt{-\text{det}\(g_{ab}+(2\pi\a')
  \tilde{F}_{ab}\)}, \no
&S_{WZ}=\frac{1}{2}N_fT_{D7}(2\pi\a')^2\int P[C_4]\wg \tilde{F}\wg \tilde{F}.
\end{align}
Here $T_{D7}$ is the D7 brane tension. $g_{ab}$ and $\tilde{F}_{ab}$ are the
  induced metric and worldvolume field strength respectively. Defining
\begin{align}
&F_{ab}=(2\pi\a')\tilde{F}_{ab}, \no
&\cN=N_fT_{D7}2\pi^2=\frac{N_fN_c}{(2\pi)^4},
\end{align}
we simplify the action to
\begin{align}\label{S_redef}
&S_{DBI}=-\frac{\cN}{2\pi^2}\int d^8\x\sqrt{-\text{det}\(g_{ab}+{F}_{ab}\)}, \no
&S_{WZ}=\frac{1}{4\pi^2}\cN\int P[C_4]\wg F\wg F.
\end{align}
The mass of the quark is realized as the separation of the D7 branes from the D3 branes at infinity. Explicitly, the mass $M$ is 
determined from the asymptotic behavior of $\th$:
\begin{align}\label{mc}
\sin\th=\frac{m}{\r}+\frac{c}{\r^3}+\cdots.
\end{align}
with $M=r_0m$.
We will turn on a constant magnetic field, which
amounts to including worldvolume magnetic field in D7 branes.
There are two possible embeddings with D7 branes crossing/not crossing
the black hole horizon, corresponding to meson melting/mesonic phase
respectively \cite{Mateos:2006nu,Hoyos:2006gb,Filev:2007gb,Erdmenger:2007bn}.
Using $t$, $\vec{x}$, $\r$ and angular coordinates on $S_3$ as
worldvolume coordinates, the induced metric is given by
\begin{align}\label{ind_metric}
ds^2_{\text{ind}}=-\frac{r_0^2}{2}\frac{f^2}{H}\r^2dt^2+\frac{r_0^2}{2}H\r^2d\vec{x}^2+\(\frac{1}{\r^2}+\th'(\r)^2\)d\r^2+\cos^2\th d\O_3^2.
\end{align}
We also turn on a constant magnetic field in
$z$-direction: $F_{xy}=B$, the action of D7 branes can be written as
\begin{align}\label{dbi_th}
S_{DBI}=-\cN\int d\r\(\frac{r_0^2}{2}\)^2fH\r^3\sqrt{1+\r^2\th'^2}\sqrt{1+\frac{2B^2}{r_0^2H\r^2}}\cos^3\th,
\end{align}
with a vanishing WZ term. The phase diagram in the $m$-$B$ plane has been obtained in \cite{Filev:2007gb,Erdmenger:2007bn}. We reproduce the result in appendix A and show the result at
fixed temperature in Figure.~\ref{fig_mB}. The two phases are mesonic phase with larger $m$ and $B$ and meson melting phase with smaller $m$ and $B$. In the former case, R-charge (axial charge) exchange between fundamental matter and adjoint sector is not possible due to the formation of meson bound state, while in the latter case, R-charge (axial charge) can leak from fundamental matter to adjoint sector. The phase diagram implies that large quark mass and magnetic field favors formation of meson bound state. The effect of magnetic field may be understood via an increased effective quark mass.
We are
interested in the meson melting phase, which is more relevant for
application in QGP.

\subsection{Fluctuations and realization of axial anomaly}

We consider the fluctuation of embedding function $\ph$ and
worldvolume gauge field $A_M$. The quadratic action can be written in
the following compact form
\begin{align}\label{compact}
S=\cN\int d^5x\(-\frac{1}{2}\sqrt{-G}G^{MN}\pd_M\ph\pd_N\ph-\frac{1}{4}\sqrt{-H}F^2\)
-\cN\k\int d^5x\Omg\e^{MNPQR}F_{MN}F_{PQ}\pd_R\ph,
\end{align}
where $M=t, x_1, x_2, x_3, \r$.
The EOM of $\ph$ is given by
\begin{align}\label{eom_ph}
\frac{\d S}{\d \ph}-\pd_M\(\frac{\d S}{\d \pd_M\ph}\)=0.
\end{align}
Since $\ph$ is a phase, only its derivative enters the action, we have
from \eqref{eom_ph},
\begin{align}
\pd_\m\(\frac{\d S}{\d \pd_\m\ph}\)+\pd_\r\(\frac{\d S}{\d \pd_\r\ph}\)=0,
\end{align}
with $\m=t, x_1, x_2, x_3$.
Defining $J_R^\m=\int d\r \frac{\d S}{\d \pd_\m\ph}$, we obtain
\begin{align}
\pd_\m J^\m_R+\frac{\d S}{\d \pd_\r\ph}\vert_{\r=\r_h}^\infty=0.
\end{align}
We will identify $J_R$ as the axial current.
The non-conservation of $J_R$
follows from two boundary terms in the integration. The boundary term
at the horizon $\r=\r_h$ indicates axial charge exchange between D7 branes and D3
branes. It is pointed out in \cite{Hoyos:2011us} that this term represents leakage
of R-charge from fundamental sector to adjoint sector as fields
in both sectors are charged under the $U(1)_R$ symmetry. The other
boundary term at $\r=\infty$ can be related to axial anomaly:
\begin{align}\label{Oph}
O_\ph\equiv -\frac{\d S}{\d \pd_\r\ph}\vert_{\r=\infty}=-\frac{\d
  S^\pd}{\d \ph(\r\to\infty)},
\end{align}
where we have used the defining property of on-shell action $S^\pd$.
For action \eqref{compact},
we have
\begin{align}\label{Oph_split}
O_\ph=\cN\sqrt{-G}G^{M\r}\pd_M\ph\vert_{\r=\infty}+\k\cN\Omg\e^{MNPQ}F_{MN}F_{PQ}\vert_{\r=\infty}.
\end{align}
For our model,
\begin{align}
&\sqrt{-G}G^{MN}=\sqrt{-h}g^{MN}g_{\ph\ph},\quad \sqrt{-H}=\sqrt{-h},
  \no
&\Omg=\cos^4\th,\quad \k=\frac{1}{8},
\end{align}
with $h$ to be defined in the next section.
Field theory analysis shows that \cite{Hoyos:2011us}\footnote{Note that we have $\ph=0$, thus no axial chemical potential is introduced.}
\begin{align}\label{Oph_O}
O_\ph=mi\bar{\psi}\g^5\psi+\cdots+\cN E\cdot B,
\end{align}
where $\cdots$ represents contribution from supersymmetric partners.
Noting that $\th\to0$ as $\r\to \infty$, we readily identify the second
term in \eqref{Oph_split} with the last term in \eqref{Oph_O}. The
remaining term in \eqref{Oph_split} can then be identified with the
mass term in \eqref{Oph_O}. For convenience, we define the remaining term by
\begin{align}\label{Oeta}
O_\eta=\cN\sqrt{-G}G^{M\r}\pd_M\ph\vert_{\r=\infty}.
\end{align}
We have thus holographically split $O_\ph$ into the mass term $O_\eta$
and anomaly term $\cN E\cdot B$, which represent respectively explicit
and anomalous breaking of axial symmetry:
\begin{align}\label{anomaly_split}
\pd_\m J^\m_R=O_\ph=O_\eta+\cN E\cdot B.
\end{align}

\section{Finite Quark Mass Effect}

We will study two aspects of finite quark mass effect:
i, the mass term, similar to QCD anomaly term, has diffusive behavior at low frequency. This gives rise to fluctuation (random walk behavior) of axial charge. The rate of diffusion, to be referred to as mass diffusion rate, determines the rate of axial charge generation;
ii, in the presence of nonvanishing $E\cdot B$, net axial charge would
be produced. However, the axial charge dissipates due to
finite quark mass, resulting in a reduced rate of axial charge
generation. 
We will refer to this as mass dissipation effect.
The above effects are captured by correlators of $J^z$ and
$O_\eta$. The mass diffusion rate involves the
correlator of $O_\eta$ itself, while
the mass dissipation effect involves the correlator between
$J^z$ and $O_\eta$. We stress that $J^z$ is the vector current
coupled to boundary gauge field $A_z$. In holographic formulation, we need to study the fluctuation of bulk fields $A_z$ and $\ph$, which are dual to $J^z$ and $O_\eta$($O_\ph$).
For our purpose, it is sufficient to turn on homogeneous (in both $\vec{x}$ and $S_3$) fluctuation of $A_z(t,\r)$ and $\ph(t,\r)$. The fluctuation leads to the following modification of the following quantities
\begin{align}\label{mod}
&ds^2_{\text{ind}}=\(g_{tt}+g_{\ph\ph}\dot{\ph}^2\)dt^2+g_{xx}dx^2+\(g_{\r\r}+g_{\th\th}\th'^2+g_{\ph\ph}\ph'^2\)d\r^2+g_{SS}^2d\O_3^2+2g_{\ph\ph}\dot{\ph}\ph'dtd\r, \no
&\d F=\dot{A}_zdt\wg dz+A_z'd\r\wg dz, \no
&\d P[C_4]=-\cos^4\th\(\dot{\ph}dt+\ph'd\r\)\wg d\O_3.
\end{align}
With \eqref{mod}, we can write down the quadratic action of $A_z(t,\r)$ and $\ph(t,\r)$:
\begin{align}\label{quadratic}
S_{\text{DBI}}+S_{\text{WZ}}&=-\cN
\int dt d^3xd\r \bigg[\frac{1}{2}\sqrt{-h}
\(g^{tt}g_{\ph\ph}\dot{\ph}^2+g^{\r\r}g_{\ph\ph}\ph'{}^2+g^{tt}g^{xx}\dot{A}_z^2+g^{\r\r}g^{xx}A_z'{}^2\) \no
&+\cos^4\th B\(\ph'\dot{A}_z-\dot{\ph}A_z'\)\bigg],
\end{align}
where we have defined
\begin{align}\label{h_def}
\sqrt{-h}=\sqrt{-g_{tt}g_{xx}\(g_{xx}^2+B^2\)\(1+\r^2\th'^2\)g_{\r\r}g_{SS}^3}.
\end{align}
Variation with respect to the fluctuations gives both the EOM and the on-shell action
\begin{align}
\d S&=-\cN\int dtd^3xd\r \bigg[\sqrt{-h}\times \no
&\(-\pd_t(g^{tt}g_{\ph\ph}\dot{\ph})\d\ph-\pd_\r(g^{\r\r}g_{\ph\ph}\ph')\d\ph-\pd_t(g^{tt}g^{xx}\dot{A}_z)\d A_z-\pd_\r(g^{\r\r}g^{xx}A_z')\d A_z\) \no
&-\pd_\r(\cos^4\th B\dot{A}_z)\d\ph+\pd_t(\cos^4\th BA_z')\d\ph-\pd_t(\cos^4\th B\ph')\d A_z+\r(\cos^4\th B\dot{\ph})\d A_z\bigg] \no
&-\cN \int dtd^3x\bigg[\sqrt{-h}\(g^{\r\r}g_{\ph\ph}\ph'\d\ph+g^{\r\r}g^{xx}A_z'\d A_z\)+\cos^4\th B(\dot{A}_z\d\ph-\dot{\ph}\d A_z)\bigg].
\end{align}
Working with a single Fourier mode $e^{-i\o t}$, we obtain the EOM
\begin{align}\label{eom_fluc}
&\o^2\sqrt{-h}g^{tt}g_{\ph\ph}\ph-\pd_\r\(\sqrt{-h}g^{\r\r}g_{\ph\ph}\ph'\)-B\pd_\r(\cos^4\th)A_z(-i\o)=0, \no
&\o^2\sqrt{-h}g^{tt}g^{xx}A_z-\pd_\r\(\sqrt{-h}g^{\r\r}g^{xx}A_z'\)+B\pd_\r(\cos^4\th)\ph(-i\o)=0.
\end{align}
The asymptotic expansion of $\ph$ and $A_z$ can be determined from EOM:
\begin{align}
&\ph=f_0+\frac{f_1}{\r^2}+\frac{f_h}{\r^2}\ln\r+\cdots, \no
&A_z=a_0+\frac{a_1}{\r^2}+\frac{a_h}{\r^2}\ln\r+\cdots.
\end{align}
$f_0$ and $a_0$ correspond to sources coupled to $O_\ph$ and $J^z$. The coefficients of the logarithmic terms correspond to counter terms\footnote{Counter terms proportional to $B^2$ can in principle exist, but are not found in this case.}:
\begin{align}
f_h=\frac{\o^2}{r_0^2}f_0,\quad a_h=\frac{\o^2}{r_0^2}a_0.
\end{align}
The vevs of  $O_\ph$ and $J^z$ are determined by
\begin{align}\label{vev}
O_\ph=&\frac{\d S^\pd}{\d \ph(\r\to\infty)} \no
=&\(-\cN\sqrt{-h}g^{\r\r}g_{\ph\ph}\ph'-\cN\cos^4\th B\dot{A}_z\)\vert_{\r\to\infty}=2\cN\(\frac{r_0^2}{2}\)^2m^2f_1-\cN Ba_0(-i\o), \no
J^z=&\frac{\d S^\pd}{\d A_z(\r\to\infty)} \no
=&\(-\cN\sqrt{-h}g^{\r\r}g^{xx}A_z'+\cN\cos^4\th B\dot{\ph}\)\vert_{\r\to\infty}=2\cN\(\frac{r_0^2}{2}\)a_1+\cN Bf_0(-i\o),
\end{align}
where we have used $\r\sin\th\vert_{\r\to\infty}= m$ according to
\eqref{mc}.
Comparing \eqref{Oeta} and \eqref{vev}, we arrive at the following dictionary
\begin{align}\label{dict}
O_\eta=2\cN\(\frac{r_0^2}{2}\)^2m^2f_1.
\end{align}

\subsection{Mass diffusion rate and susceptibility}

The mass operator $O_\eta$ can lead to
fluctuation of axial charge. It is well known that the origin of
axial charge fluctuation from QCD anomaly is
topological transitions. The counterpart for $O_\eta$ is helicity
flipping from elementary scattering \cite{Manuel:2015zpa,Grabowska:2014efa}. The rate of axial charge
generation in case of topological transition is given by CS
diffusion rate. Similarly, the corresponding rate in case of
helicity flipping is given by the diffusion rate of $O_\eta$, which we
calculate below.
The diffusion rate of $O_\eta$ is encoded in the low frequency limit
of retarded correlator. To calculate the retarded correlator, we need
to turn on source for $\ph$ while keeping $A_z$ vanish on the boundary.
Both $\ph$ and $A_z$ satisfy infalling wave condition on the horizon.
It follows from \eqref{dict} that the retarded correlator is given by
\begin{align}\label{eta_retarded}
G_{\eta\eta}(\o)=\int dt\<[O_\eta(t),O_\eta(0)]\>\Theta(t)e^{i\o t}=-2\cN\(\frac{r_0^2}{2}\)^2m^2\frac{f_1}{f_0}.
\end{align}
The diffusion rate is defined by
\begin{align}\label{Gm_def}
\G_m=\lim_{\o\to0}\frac{2iT}{\o}G_{\eta\eta}(\o).
\end{align}
For the case $B=0$, there is no mixing between $\ph$ and $A_z$.
We can simply use $\ph^{(0)}$ \eqref{sol_homo}
in appendix B:
\begin{align}
&\ph^{(0)}(\r)=1-\frac{i\o}{2r_0}\big[\int_1^\r
  d\r' \(\frac{\(\frac{r_0^2}{2}\)^28\cos^3\th_h\sin^2\th_h}{\sqrt{-h(\r')}g^{\r\r}(\r')g_{\ph\ph}(\r')}-\frac{1}{\r'-1}\)+\ln(\r-1)\big].
\end{align}
This gives the following retarded correlator of $O_\eta$:
\begin{align}\label{Oeta_retarded}
G_{\eta\eta}(\o)=-2\cN\(\frac{r_0^2}{2}\)^2\frac{i\o}{4r_0}8\cos^3\th_h\sin^2\th_h.
\end{align}
\eqref{Oeta_retarded} gives a mass diffusion rate $\G_m$ as analog of CS diffusion rate:
\begin{align}\label{Gamma_m}
\G_m=\frac{\cN}{\pi}\(\frac{r_0^2}{2}\)^28\cos^3\th_h\sin^2\th_h.
\end{align}
The dependence on $m$ is encoded in the combination of
trigonometric functions, which clearly indicates an upper bound of the
mass diffusion rate. We also extract $\G_m$ using \eqref{Gm_def} with
numerical solutions for general $B$ and $m$ in the meson melting phase.
\begin{figure}[t]
\includegraphics[width=0.5\textwidth]{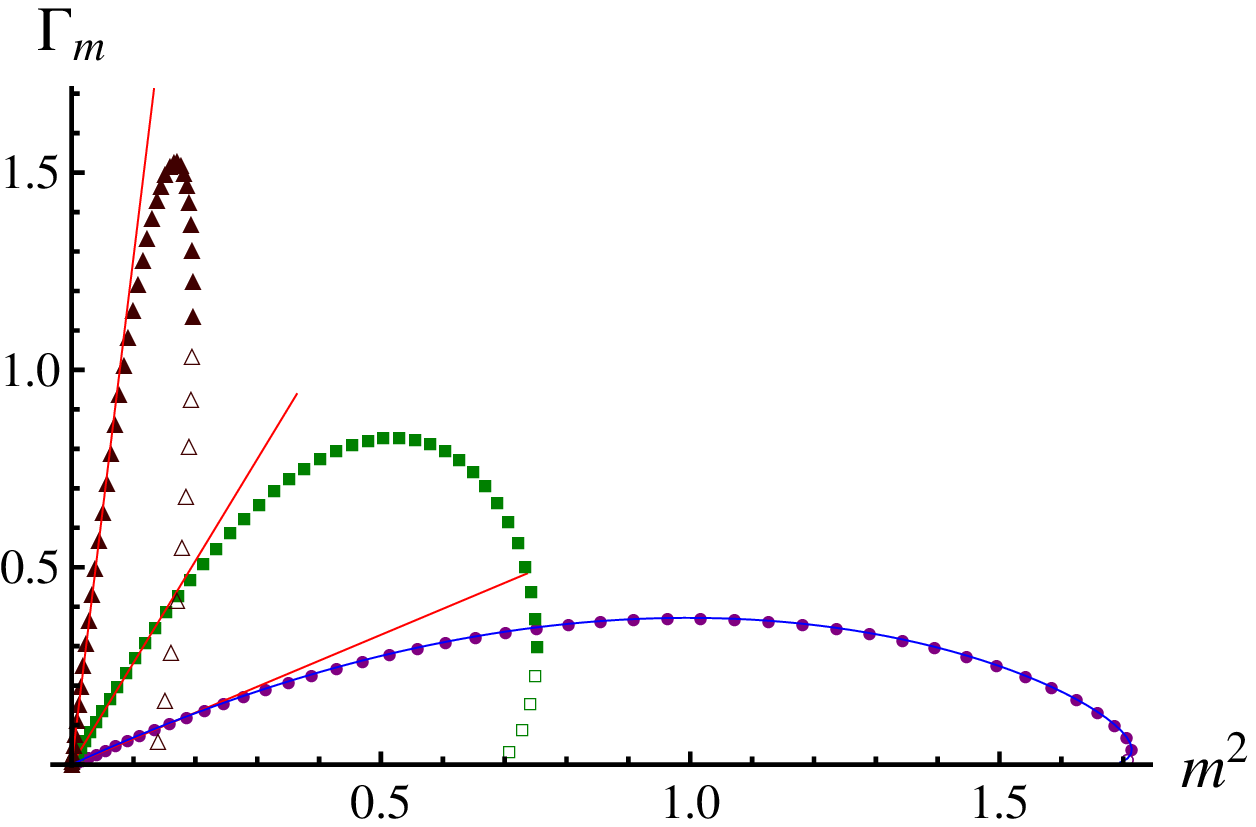}
\includegraphics[width=0.5\textwidth]{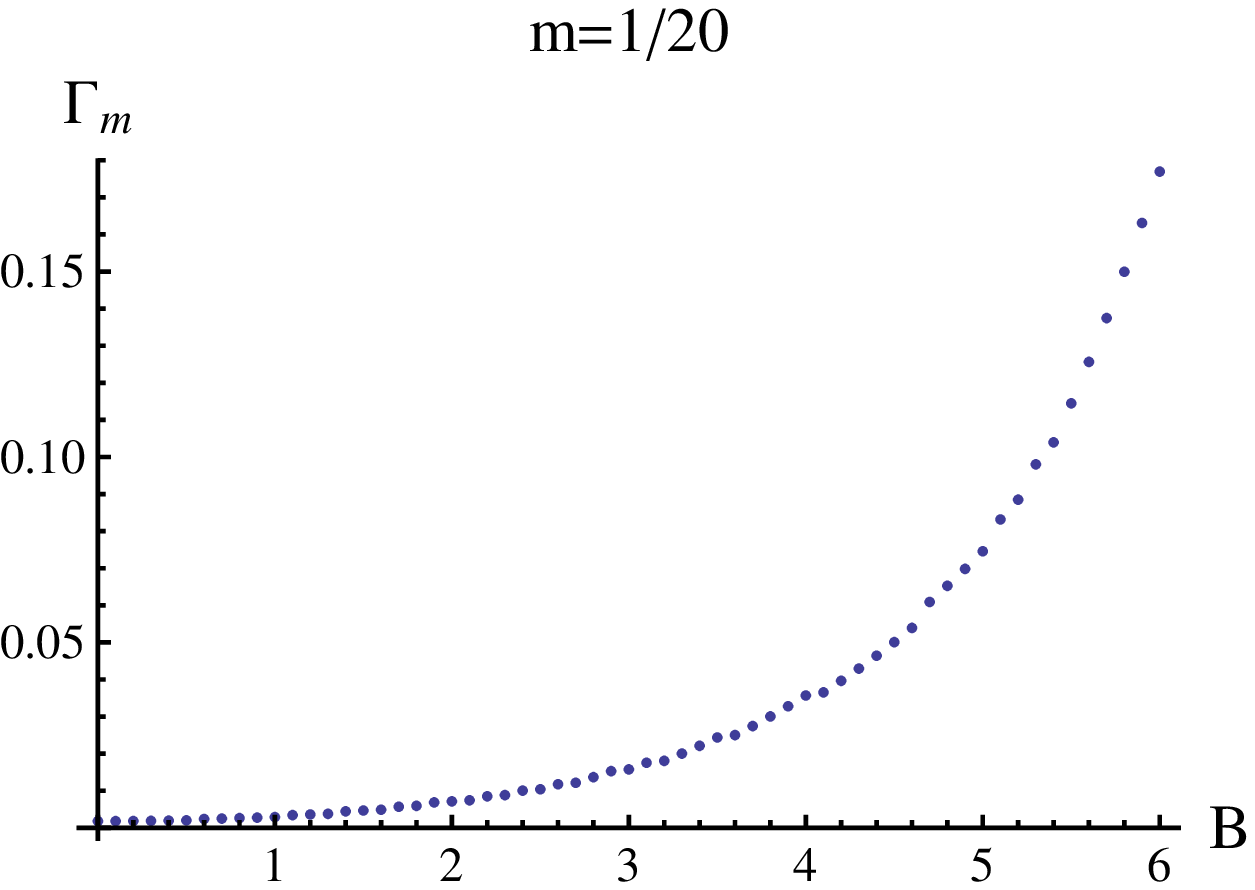}
\caption{\label{fig_gamm}(left) The mass diffusion rate ${\Gamma}_m$ as a function of $m^2$ for $B=0$ (blue point), $B=2$ (purple square), $B=4$ (brown diamond). The units are set by $\pi T=1$. The blue line is given by \eqref{Gamma_m} which fits well for $B=0$. To guide eyes, we also include linear fittings (red line) in the small mass region. The linear behavior is consistent with field theory expectation. We have used empty symbols for points in metastable phases. (right) $\G_m$ as a function of $B$ at $m=1/20$. A rapid growth of $\G_m$ with $B$ is found.}
\end{figure}
We plot numerical results of $\G_m$ as a function
of $m^2$ for different values of $B$ in
Figure.~\ref{fig_gamm}. The case $B=0$ agrees well with analytic
expression \eqref{Oeta_retarded}. We find the mass diffusion rate is a
non-monotonous function of $m$. This is not difficult to understand:
in the limit $m\to0$, $\G_m$ obviously should vanish as $O_\eta\sim
m$. When $m$ approaches the phase boundary between meson melting phase
and mesonic phase, we also expect helicity flipping to freeze due to
formation of meson bound states. In between, there must be a maximum
for $\G_m$. Furthermore, the linear behavior of $\G_m$-$m^2$ plot in
small $m$ region supports the scaling $\G_m\sim m^2$, which is
consistent with field theory
expectation. 
The $B$ dependence is more interesting: $\G_m$ shows rapid growth with $B$.
The presence of $B$ enhances
the diffusion, which cannot be explained as the increase of effective
mass. 
The enhancement of helicity flipping might provide a way to
generate axial charge more efficiently. It is worth mentioning that an enhancement of CS
diffusion rate due to magnetic field was also obtained in \cite{Basar:2012gh,Drwenski:2015sha}.

We would like to comment on the diffusive behavior of $O_\eta$. On general ground, the mass diffusion effect leads to accumulation of axial charge, which prevents its further generation. It would lead to modification of the long time (low frequency) behavior of $G^R_{\eta\eta}$. However,
this does not happen due to the existence of the adjoint reservoir. The generated axial charge entirely dissipates to the adjoint reservoir. To see that, we compare $O_\eta$ and $O_\text{loss}$, which are the same quantity below evaluated at $\r=\infty$ and $\r=1$ respectively.
\begin{align}
\cN\sqrt{-h}g^{\r\r}\r_{\ph\ph}\ph'.
\end{align}
It follows from the EOM \eqref{eom_fluc} that the above quantity is constant in the limit $\o\to 0$,
\begin{align}
\pd_\r\(\cN\sqrt{-h}g^{\r\r}\r_{\ph\ph}\ph'\)=0,
\end{align}
meaning that the generated charge is entirely balanced by the loss to the reservoir.
Consequently, the low frequency behavior of $O_\eta$ correlator is still diffusive.

Turing on the source for $O_\eta$ also allows us to study the susceptibility of axial charge. In the presence of finite quark mass, the axial charge is not even approximately conserved, making the susceptibility a subtle concept. Following \cite{Iatrakis:2015fma}, we can use CME to define a dynamical susceptibility $\c$. In the present model, it is given by
\begin{align}\label{chi}
\c=\frac{n_5}{\mu_5}=\frac{\cN B n_5}{J^z}.
\end{align}
We need to calculate both $n_5$ and $J^z$ from response to source for $O_\ph$ in the hydrodynamic limit. $n_5$ is essentially known already. Denoting the source by $f_m$, we can express $n_5$ as
\begin{align}
-i\o n_5(\o)=O_\eta(\o)=-G^R_{\eta\eta}(\o)f_m(\o)\sim O(\o)f_m(\o).
\end{align}
Therefore we obtain $n_5\sim O(\o^0)f_m(\o)$. On the other hand, $J^z$ is calculated using the dictionary \eqref{vev}. It is generated through the mixing between $\ph$ and $A_z$. $J^z$ is also expressible as response to $f_m$
\begin{align}
J^z(\o)=-G^R_{j\eta}(\o)f_m(\o).
\end{align}
Using the hydrodynamic solution \eqref{A1} in appendix B and the dictionary \eqref{vev}, we find that there are two contributions to $G^R_{j\eta}$, both of which are of order $O(\o B)$. Therefore we have $J^z\sim O(\o B)f_m(\o)$. Plugging the above qualitative results into \eqref{chi}, we obtain
\begin{align}
\c\sim O(\o^{-1}).
\end{align}
It simply means that the susceptibility is divergent in the static limit $\o\to 0$. Recalling that the susceptibility is well-defined in the massless limit, we arrive at the non-commutativity of the limits $m\to0$ and $\o\to0$. The physical reason for divergent susceptibility is not difficult to understand. On on hand, the mass diffusion effect can spontaneously generate axial charge density at the cost of no energy. On the other hand, as we have seen already, the adjoint reservoir is a perfect sink for axial charge in the flavor sector, preventing accumulation of axial charge. Consequently, the axial charge can be continuously generated in the flavor sector.
Note that the situation is different in case of axial charge generation by QCD anomaly. There the breaking of axial symmetry is suppressed by $1/N_c$ (or the quenched limit), resulting in a finite dynamical susceptibility.

\subsection{Mass dissipation effect}

\begin{figure}[t]
\includegraphics[width=0.5\textwidth]{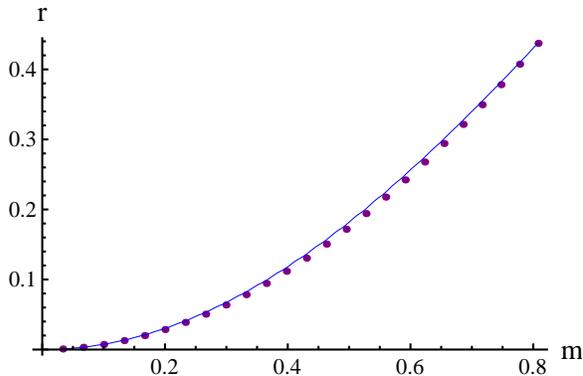}
\caption{\label{fig_Oetam}The mass dissipation rate $r$ as a function of $m$ from numerics with small $B$ and small $\o$. It is a monotonous increasing function of $m$ as expected. The analytic function \eqref{r_analytic} is drawn in blue line and fits the numerical results well. The units are set by $\pi T=1$.}
\end{figure}

\begin{figure}[t]
\includegraphics[width=0.5\textwidth]{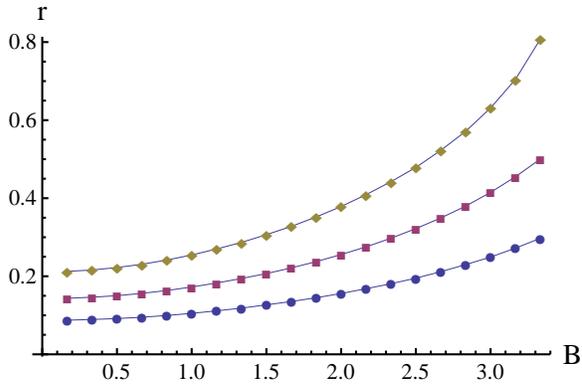}
\caption{\label{fig_OetaB}The mass dissipation rate $r$ as a function of $B$ for $m=7/20$ (blue point), $m=9/20$ (purple square), and $m=11/20$ (brown diamond). The units are set by $\pi T=1$. The analytic function \eqref{r_analytic} is drawn in blue line and fits the numerical results well. It is a monotonous increasing function of $B$.}
\end{figure}

To study the mass dissipation effect, we turn on an electric field in
$z$-direction by a time dependent $A_z$ on the boundary. We do not
need to source $\ph$ on the boundary. Its profile is entirely
generated via mixing of $A_z$ and $\ph$ in the bulk. The resulting
$O_\eta$ from nontrivial profile of $\ph$ corresponds to the mass
dissipation effect we are after. We also impose infalling wave boundary
condition for $\ph$ and $A_z$ since we are interested in calculating response.
We define the dimensionless mass dissipation rate
\begin{align}\label{mir}
r=\frac{O_\eta}{\cN E\cdot B}.
\end{align}
The rate is a function of $\o$, $m$ and $B$. In the hydrodynamic
limit $\o\to 0$, we can show that $r(\o\to 0)$ is a real function of $m$
and $B$. In fact it can be related to embedding function for given $m$ and $B$ in the
meson melting phase. To obtain $r(\o\to 0)$ analytically, we need to solve the
coupled EOM \eqref{eom_fluc} in the hydrodynamic limit. The hydrodynamic
solutions can be found in appendix B. We simply quote the results here.
The leading
nontrivial order is the zeroth order for $A_z$ and the first order for $\ph$:
\begin{align}\label{hydro_sol}
&A_z^{(0)}=a_0, \no
&\ph^{(1)}=\frac{\(1-\cos^4\th_h\)B i\o a_0}{\(\frac{r_0^2}{2}\)^2m^2(-2)}\r^{-2}+\cdots,
\end{align}
where $\th_h$ is the value of $\th$ on the horizon, which needs to be obtained from numerical embedding function for given $m$ and $B$.
For $\ph^{(1)}$, we only retain its asymptotic behavior relevant for
extracting $O_\eta$. \eqref{hydro_sol} leads to the following rate
\begin{align}\label{r_analytic}
r=1-\cos^4\th_h.
\end{align}
We also study the rate of dissipation by numerical solutions. In practice,
we generate two independent infalling numerical solutions at the horizon and use their linear combination to construct the solution with desired boundary condition. 
We show $m$-dependence of $r$ in the limit $B=0$ and $B$-dependence of $r$ at
different values of $m$ in Figure~\ref{fig_Oetam} and
Figure~\ref{fig_OetaB}. We find good agreement with analytic expression \eqref{r_analytic}.
On general ground, we expect the rate to be a monotonous increasing function of
$m$. In particular, $r\to 0$ as $m\to0$. Indeed, this is confirmed in Fig.~\ref{fig_Oetam}.
We further note that the effect of $B$ enhances the
dissipation on top of the mass effect in Fig.~\ref{fig_OetaB}. 
The physical interpretation of the dissipation rate $r$ turns out to be a subtle question. Recalling the axial anomaly equation \eqref{anomaly_split}, we would draw the following conclusion: for every one unit of axial charge generated by parallel electric and magnetic field, $r$ unit of it dissipates through the mass term, with a unit of $1-r$ axial charge remaining. The remaining axial charge survives even in the hydrodynamic limit since we have $\o\to 0$. This is not true because we have ignored a third source of axial charge dissipation, i.e. loss to the adjoint reservoir. The anomaly equation \eqref{anomaly_split} should be supplemented by the loss rate 
\begin{align}\label{anomaly_complete}
\pd_\m J^\m_R=O_\eta+\cN E\cdot B-O_\text{loss},
\end{align}
with the explicit form of loss rate given by
\begin{align}\label{loss}
O_\text{loss}=\cN\sqrt{-h}g^{\r\r}r_{\ph\ph}\pd_\r\ph\vert_{\r\to 1}+\cN\O E\cdot B\vert_{\r\to 1}.
\end{align}
It is known that the loss rate can be IR unsafe \cite{Karch:2008uy}. Indeed, plugging in the hydrodynamic solution $A_z^{(0)}$ and $\ph^{(1)}$ in appendix B into \eqref{anomaly_complete}, we find both terms becomes infinitely oscillatory as $\r\to1$. Nevertheless we can still extract useful information by taking the hydrodynamic limit $\o\to0$ before the IR limit $\r\to 1$. Using this regularization we find the $\cN\O E\cdot B$ term becomes $\cN \cos^4\th_h E\cdot B$, while the other term is higher order in $\o$. We immediately note that $\cN \cos^4\th_h E\cdot B$ is precisely the $1-r$ unit of axial charge. Subtracting the charge loss to the reservoir, we find only $r$ unit of axial charge is effectively generated in the flavor sector by parallel electric and magnetic field. All dissipates by the mass term. This simply means no axial charge survives in the hydrodynamic limit.
After clarifying the role of axial charge loss to adjoint reservoir, we should interpret $r$ as a measure of mass dissipation effect compared to dissipation to the adjoint reservoir. The dissipation through mass term is favored at large $m$ and $B$.

The statement on the non-survival of axial charge can receive correction higher order in $\o$, which quantifies the charge survival rate.
We can compare with the relaxation time approximation employed in \cite{Jimenez-Alba:2015awa,Sun:2016gpy}, in which the following form of axial anomaly equation is assumed (here we use $r\cN E\cdot B$ for effectively axial charge generation)
\begin{align}\label{rta}
\pd_tn_5=-\frac{n_5}{\t}+r\cN E\cdot B,
\end{align}
with $\t$ being the relaxation time.
Physically it means the presence of axial charge $n_5$ induces $O_\eta=-\frac{n_5}{\t}$. Plugging it into \eqref{rta}, we can solve for $O_\eta$ in frequency space
\begin{align}\label{Oeta_rta}
O_\eta=-\frac{r\cN E\cdot B}{1-i\o\t}.
\end{align}
The leading order result $O_\eta=-r\cN E\cdot B$ corresponds to our result of full dissipation. In principle, by going to high order in $\o$, we could calculate the relaxation time $\t$. We will not attempt it in this paper.

\section{Summary}

We have investigated the effect of finite quark mass and magnetic field in the generation
and dissipation of axial charge, using a D3/D7 model.
For axial charge
generation, we calculated the mass diffusion rate. It is analogous
to the Chern-Simon diffusion rate as a measure of axial charge
fluctuation. The mass diffusion rate is a bounded
non-monotonous function of mass at vanishing magnetic field. The
presence of magnetic field enhances
the diffusion. At small $m$, our
numerical results are consistent with an approximate scaling for the mass
diffusion rate
\begin{align}
\G_m\sim m^2F(B),
\end{align}
with $F(B)$ a rapid growing function in the meson melting phase.
We also defined a dynamical susceptibility of axial charge using CME. We found the susceptibility to be divergent in the static limit $\o\to0$. It is due to two reasons: i, spontaneous generation of axial charge by mass diffusion effect; ii, continuous leakage of axial charge from flavor sector to the adjoint sector, preventing the accumulation of axial charge.

For axial charge dissipation, we found that a mass term is induced in the presence of parallel electric and magnetic fields, reducing the generation of axial charge. After carefully subtracting the axial charge loss rate to the adjoint sector, we found that the axial charge dissipate entirely through the mass term in the long time limit. To the order we consider, it is consistent with a relaxation time approximation.

\section{Acknowledgments}

We thank K.~Landsteiner and Y.~Yin for critical comments on an early version of the paper. We also thank D.~Kharzeev, J.~F.~Liao, Y.~ Liu, L.~Yaffe, H.-U.~Yee and Y.~Yin for useful discussions. 
The work of S.L. is in part supported by Junior Faculty's Fund at Sun Yat-Sen University.

\appendix

\section{Phase diagram at finite $m$ and $B$}

It is known that D3/D7 brane system as a model for finite temperature
QGP in the quenched limit has a first order phase transition
\cite{Mateos:2006nu,Hoyos:2006gb,Filev:2007gb,Erdmenger:2007bn}. At large quark
mass and strong magnetic field (with temperature fixed), the probe D7 brane lie outside of the black hole horizon. In this phase the meson stays in bound state and its spectrum possesses a mass gap. At small quark mass and weak magnetic field, the brane crosses the horizon and this corresponds to the meson melting phase.
The case in which the D7 branes touch the horizon corresponds to critical embedding, giving rise to critical mass and condensate. The embeddings close to the critical embedding show oscillatory behavior for the corresponding mass and condensate parameters around their critical values. This implies that the condensate is a multivalued function of mass, corresponding to different states. The true ground state is determined by the embedding that minimizes the free energy.
Denoting $\c=\sin\th$, we can rewrite the action \eqref{dbi_th} as
\begin{align}\label{dbi_chi}
S_{DBI}=-\cN\int d\r\frac{(1-\r^4)(1-\c^2)\sqrt{(1-\c^2+\r^2\c'^2)\(1+(2+4B^2)\r^4+\r^8\)}}{4\r^5},
\end{align}
where we have set $r_0=1$, which amounts to fixing the temperature $T=\frac{1}{\pi}$. The EOM following from \eqref{dbi_chi} is solved by numerical integration of the EOM.
The black hole embedding and Minkowski embedding satisfy different
boundary conditions. For black hole embedding, the boundary condition
is $\c(\r=1)=\c_0$, $\c'(\r=1)=0$, with integration domain from
$\r=1$ to $\r=\r_{\text{max}}$. For Minkowski embedding, the boundary
condition is $\c(\r=\r_{\text{min}})=1$,
$\c'(\r=\r_{\text{min}})=\frac{(1-\r^4)\(1+(2+4B^2)\r^4+\r^8\)}{\r(1+\r^4)(1+2B^2\r^4+\r^8)}$,
with integration domain from $\r=\r_{\text{min}}>1$ to $\r=\r_{\text{max}}$. The
initial condition for the derivative is chosen such that
$\c''(\r=\r_{\text{min}})$ can be uniquely determined by EOM. In
practice, we start the integration at $\r=1+\e$ for black hole embedding and $\r=\r_{\text{min}}+\e$ for Minkowski embedding.

\begin{figure}[t]
\includegraphics[width=0.5\textwidth]{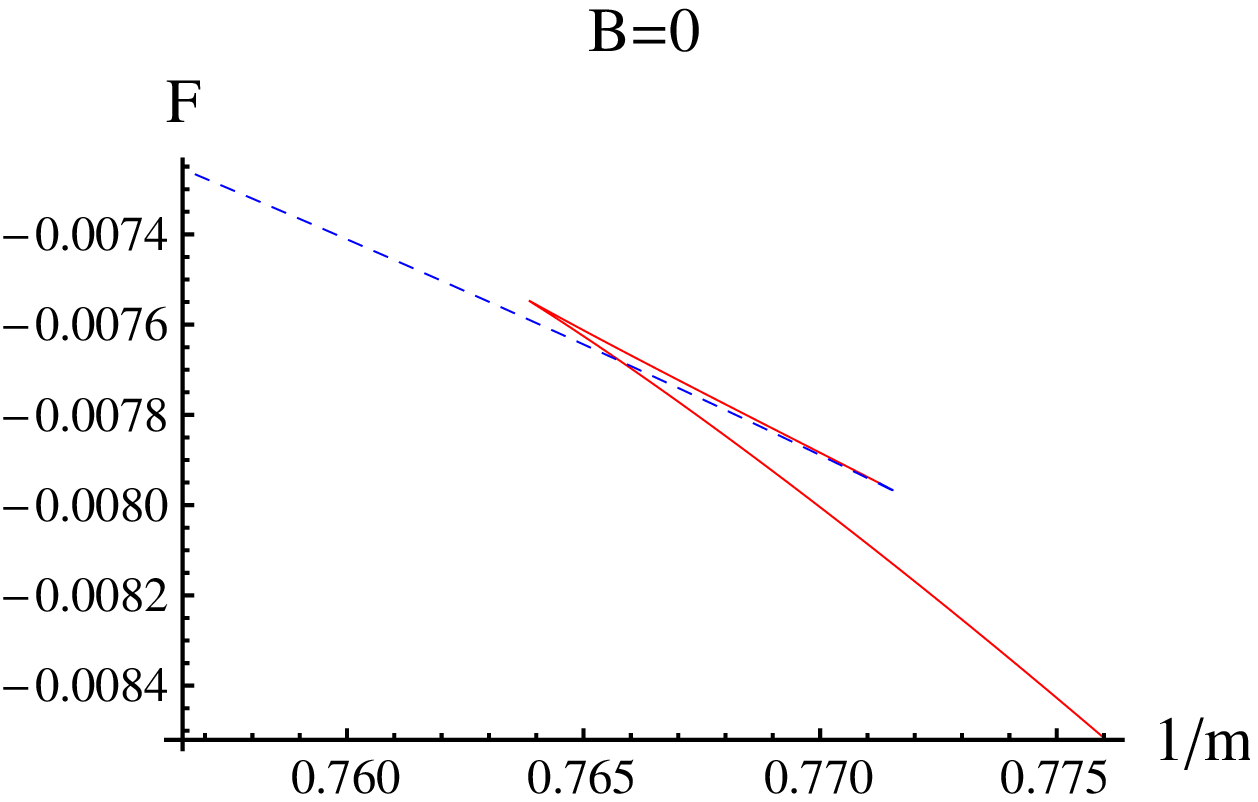}
\includegraphics[width=0.5\textwidth]{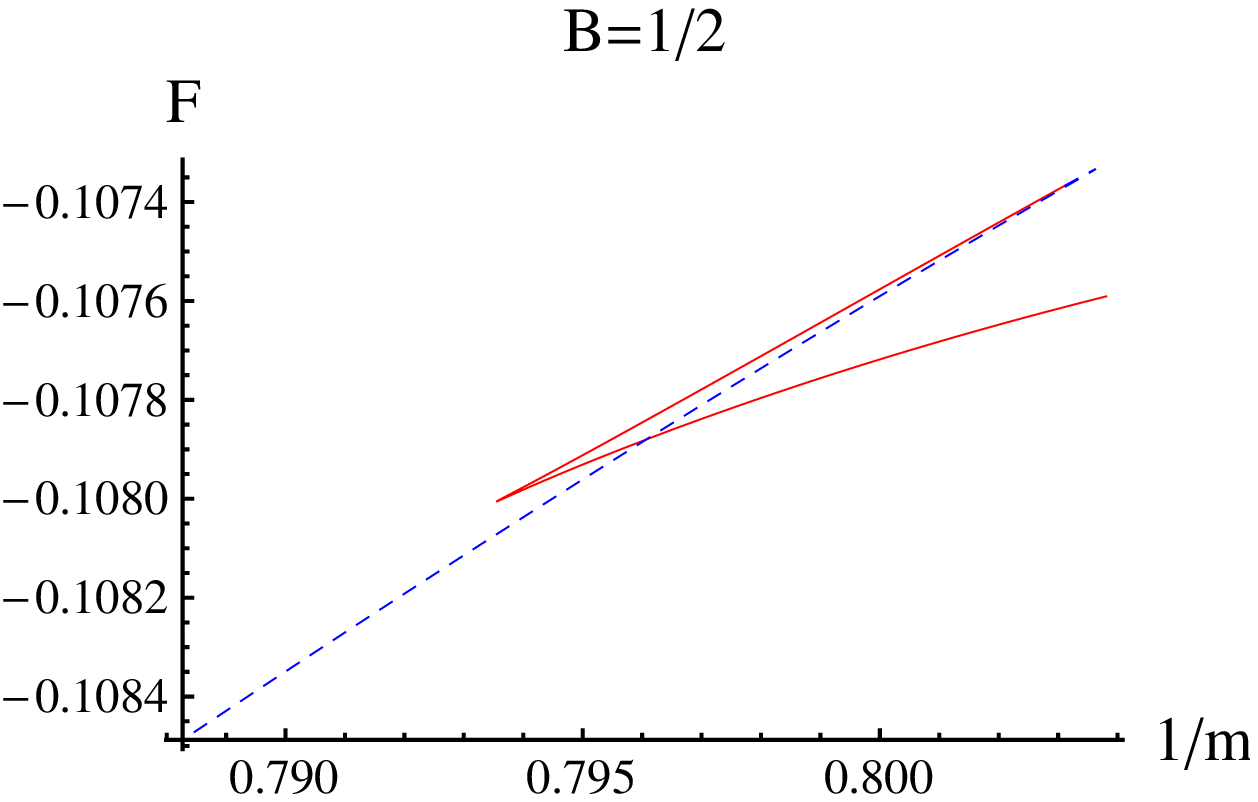}
\includegraphics[width=0.5\textwidth]{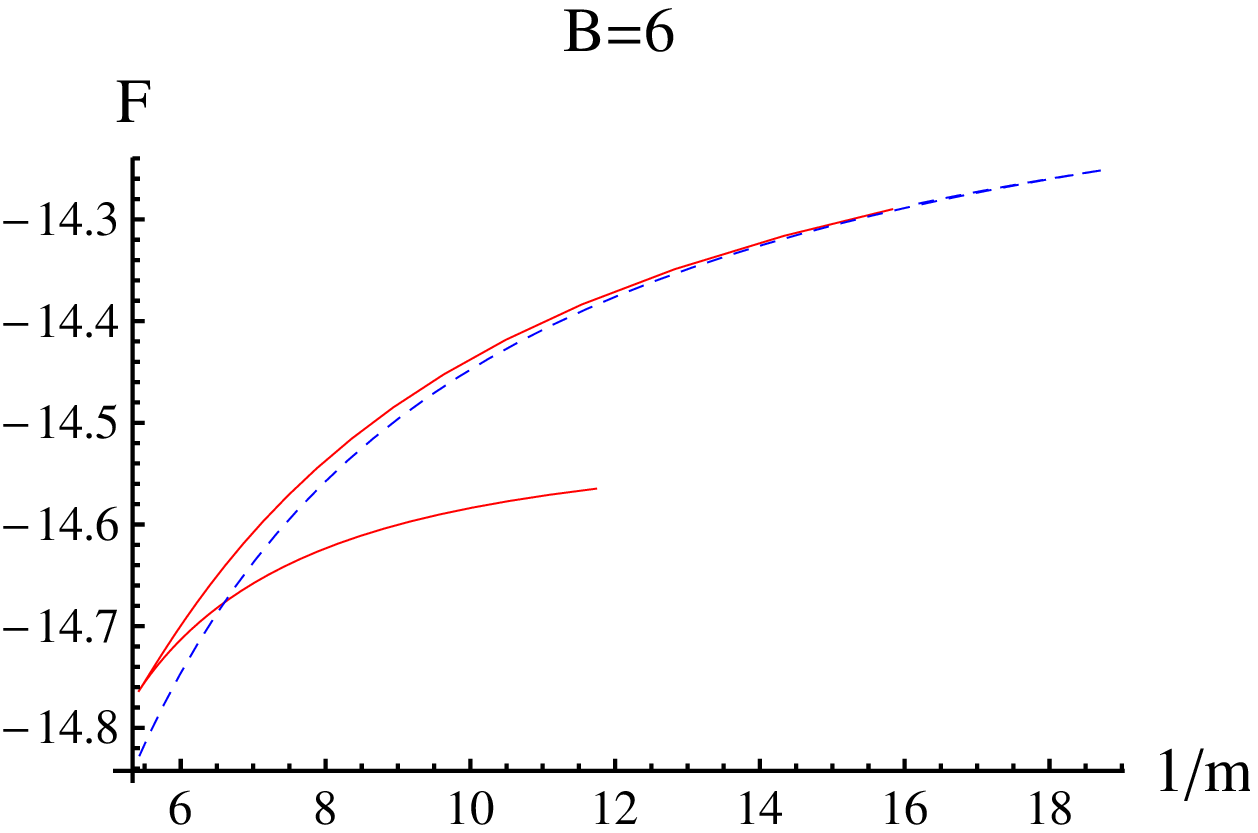}
\includegraphics[width=0.5\textwidth]{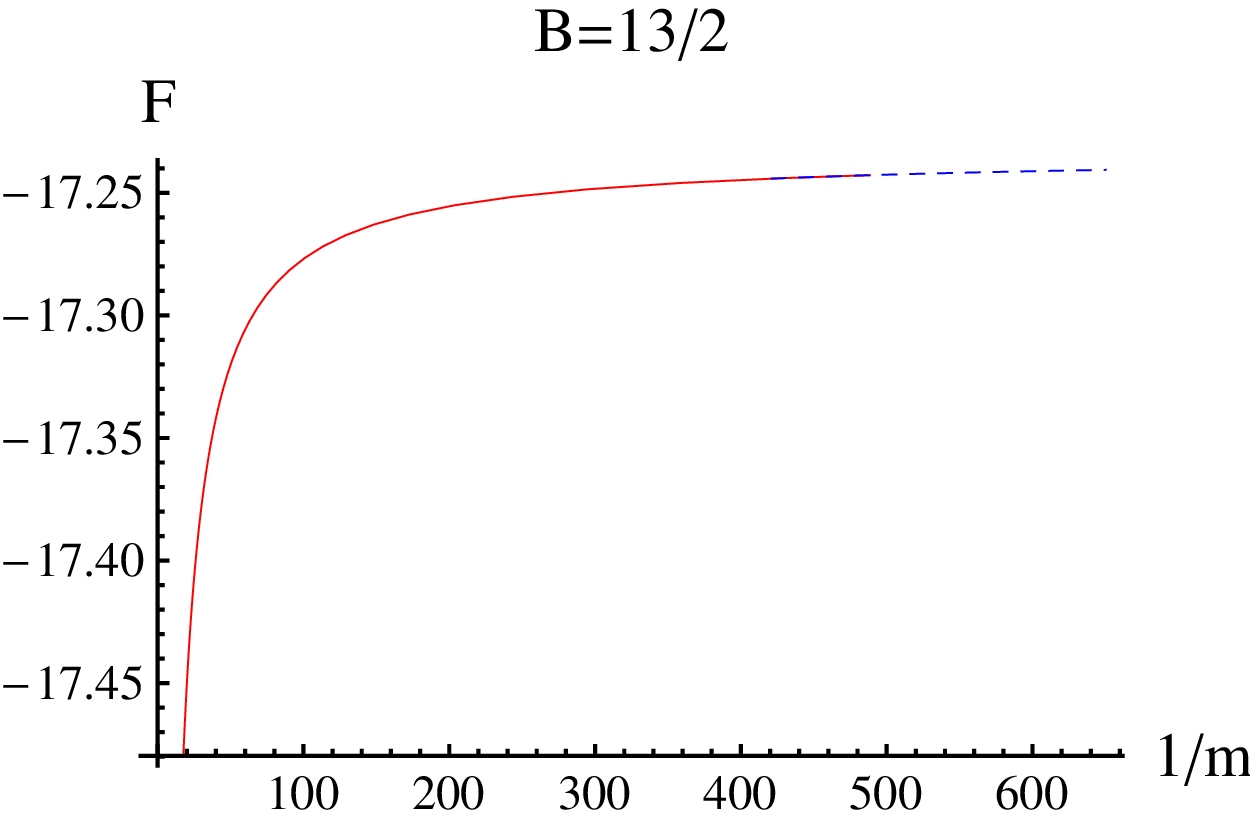}
\caption{\label{fig_Fm}Free energy $F$ as a function of $1/m$ at
  different $B$ for $D3/D7$ system. The units are set by $\pi T=1$. The red continuous (blue dashed) curves correspond to the black hole (Minkowski) embedding.}
\end{figure}

We note that the free energy $F = TS$ contains a UV divergence and
therefore needs to be renormalized. Following~\cite{Mateos:2006nu}, we add to the
action the counter term
\begin{align}\label{Counter-term}
S_{{counter}}\,=\,-\frac{\cN}{4}\big[\((\rho_{\text{max}}^2-m^2)^2-4mc\)+
\frac{B^2}{2}\ln\rho_{max}\big].
\end{align}
Note the appearance of a new term due to magnetic field as compared to \cite{Mateos:2006nu}.
The renormalized action $S = S_{DBI} + S_{{counter}}$ is finite as we
take $\rho_{\text{max}} \to \infty$. The true ground state is found by
comparing the free energy of black hole embedding and Minkowski
embedding, corresponding to meson melting phase and mesonic phase. The
phase transition is first order and present below a certain critical
magnetic field $B_c$. Above $B_c$, only Minkowski embedding is
possible. Below $B_c$, metastable phases of black hole embedding are found
as we increases $B$. We illustrate the structure of metastable phases
in Figure~\ref{fig_Fm}.

\section{Hydrodynamic solution of fluctuations}

We wish to solve \eqref{eom_fluc} in the hydrodynamic limit. We reproduce \eqref{eom_fluc} for convenience.
\begin{align}\label{eom_app}
&\o^2\sqrt{-h}g^{tt}g_{\ph\ph}\ph-\pd_\r\(\sqrt{-h}g^{\r\r}g_{\ph\ph}\ph'\)-B\pd_\r(\cos^4\th)A_z(-i\o)=0, \no
&\o^2\sqrt{-h}g^{tt}g^{xx}A_z-\pd_\r\(\sqrt{-h}g^{\r\r}g^{xx}A_z'\)+B\pd_\r(\cos^4\th)\ph(-i\o)=0.
\end{align}
For pedagogical reason, we work in the small $B$ limit and solve
\eqref{eom_app} order by order in $B$. Since correction to $\sqrt{-h}$
starts from $O(B^2)$, we can simply use the $B=0$ limit of $\sqrt{-h}$
for the solution up to $O(B)$.
The order $O(B^0)$ solution
satisfies homogeneous equation. In the hydrodynamic regime, the
solution is given by
\begin{align}\label{sol_homo}
&\ph^{(0)}(\r)=1-\frac{i\o}{2r_0}\big[\int_1^\r
  d\r' \(\frac{\(\frac{r_0^2}{2}\)^28\cos^3\th_h\sin^2\th_h}{\sqrt{-h(\r')}g^{\r\r}(\r')g_{\ph\ph}(\r')}-\frac{1}{\r'-1}\)+\ln(\r-1)\big],
  \no
&A_z^{(0)}(\r)=1-\frac{i\o}{2r_0}\big[\int_1^\r
  d\r' \(\frac{\frac{r_0^2}{2}4\cos^3\th_h}{\sqrt{-h(\r')}g^{\r\r}(\r')g^{xx}(\r')}-\frac{1}{\r'-1}\)+\ln(\r-1)\big].
\end{align}
We have chosen a specific normalization for the homogeneous solutions.
At order $O(B)$, we need to solve the inhomogeneous equations sourced
by the mixing terms. This can be achieved by using Green's function
for $\ph$ and $A_z$, which are defined by
\begin{align}\label{green}
&\pd_\r^2\cG_\ph(\r,\r')+\pd_\r
 \cG_\ph(\r,\r')\pd_\r\ln\(\sqrt{-h}g^{\r\r}g_{\ph\ph}\)-\frac{\o^2g^{tt}}{g^{\r\r}}\cG_\ph(\r,\r')=\d(\r-\r'),
  \no
&\pd_\r^2\cG_A(\r,\r')+\pd_\r
  \cG_A(\r,\r')\pd_\r\ln\(\sqrt{-h}g^{\r\r}g^{xx}\)-\frac{\o^2g^{tt}}{g^{\r\r}}\cG_A(\r,\r')=\d(\r-\r').
\end{align}
We require that the inhomogeneous solutions satisfy the infalling wave
condition on the horizon and vanish on the boundary. It is convenient
to construct the Green's function using two independent solutions
satisfying the above boundary conditions. We illustrate the procedure
using $\cG_\ph$ as an example. The two independent solutions are chosen
as below:
\begin{align}\label{ph_hb}
&\ph_h=(\r-1)^{-\frac{i\o}{2r_0}}\(1+\cdots\)=h_0+\frac{h_1}{\r^2}+\cdots,
  \no
&\ph_b=\ph_hh_0^*-\ph_h^*h_0=\(h_0^*h_1-h_0h_1^*\)\r^{-2}+\cdots.
\end{align}
Here $\ph_h$ satisfies the infalling wave condition on the horizon.
$h_0$ and $h_1$ (not to be confused with $h$) are coefficients of
asymptotic expansion of $\ph_h$.
$\ph_b$ is constructed from linear combination of $\ph_h$ and its
complex conjugate such that it vanishes on the boundary. $\cG_\ph$ can be constructed as follows
\begin{align}\label{G_ph}
\cG_\ph(\r,\r')=\frac{1}{\ph'_b(\r')\ph_h(\r') \ph_b(\r')\ph_h'(\r')}\big[\ph_h(\r')\ph_b(\r)\th(\r-\r')+\ph_b(\r')\ph_h(\r)\th(\r'-\r)\big].
\end{align}
The Wronskian appearing in \eqref{G_ph} can be fixed up to
normalization from the homogeneous equation:
\begin{align}\label{wronskian}
\ph_b'\ph_h-\ph_h'\ph_b=\frac{\#}{\sqrt{-h}g^{\r\r}g_{\ph\ph}}.
\end{align}
We can fixed the normalization by taking the limit $\r\to\infty$ of
\eqref{wronskian}. Comparing the limit with \eqref{ph_hb}, we obtain
\begin{align}
\#=\(\frac{r_0^2}{2}\)^2m^2\(h_0^*h_1-h_0h_1^*\)(-2)h_0,
\end{align}
where we used the fact $\r\sin\th\vert_{\r\to\infty}=m$.
The inhomogeneous solution is given by the convolution of Green's function and
corresponding source
\begin{align}\label{ph_inhomo}
\ph ^{(1)} (\r)=\int_1^\infty d\r'\cG_\ph(\r,\r')s(\r'),
\end{align}
with the source
\begin{align}
s(\r)=\frac{B\pd_\r(\cos^4\th)A_z^{(0)}i\o}{\sqrt{-h}g^{\r\r}g_{\ph\ph}}.
\end{align}
We are only interested in the limit $\r\to\infty$ of
\eqref{ph_inhomo}, which is
\begin{align}\label{ph1}
\ph ^{(1)} (\r)=\int_1^\infty d\r'\frac{\ph_h(\r')B\pd_{\r'}(\cos^4\th(\r'))A_z^{(0)}(\r')i\o}{\(\frac{r_0^2}{2}\)^2m^2(-2)h_0}\r^{-2}+\cdots.
\end{align}
Following the same procedure, we obtain the counterpart of $A_z$:
\begin{align}\label{A1}
A_z^{(1)}(\r)=\int_1^\infty d\r'\frac{-A_{z,h}(\r')B\pd_{\r'}(\cos^4\th(\r'))\ph^{(0)}(\r')i\o}{\(\frac{r_0^2}{2}(-2)a_0\)}\r^{-2}+\cdots,
\end{align}
where $A_{z,h}$ is defined as the solution satisfying infalling wave condition on the horizon, with boundary value $a_0$.
Before closing this section, we claim that \eqref{sol_homo}, \eqref{ph1}
and \eqref{A1} can also be understood as series expansions in $\o$: we
should discard the $O(\o)$ terms in \eqref{sol_homo} and view the rest
as zeroth solution. The first order solutions get contribution from
the discarded terms in \eqref{sol_homo}, \eqref{ph1} and
\eqref{A1}. Furthermore, we can allow for arbitrary dependence of $\sqrt{-h}$ on $B$ in \eqref{sol_homo}, \eqref{ph1} and \eqref{A1} provided that
we work with sufficient small $\o$.

\bibliographystyle{unsrt}
\bibliography{Q5ref}

\end{document}